\input{aipcheck}

\documentclass[
    ,final            
  ]
  {aipproc}

\layoutstyle{6x9}

\begin{document}

\title{Coherent Pion Production by Neutrinos}


\author{E.A.~Paschos}{
  address={University of Dortmund, Institute of Physics, D--44221
           Dortmund, Germany}
}



\begin{abstract}
I concentrate in this article on the reaction coherent pion production
by neutrinos incident on nuclei.  A special effort is made in order to
describe the approximations entering the calculation.  I conclude that
the reaction is well understood and with appropriate data for hadronic
reactions it can be computed for low and high neutrino energies. Because
of shortage of space I omitted the resonance analysis, which is described
in articles with my collaborators.
\end{abstract}

\maketitle





During the past few years coherent pion production by neutrinos 
attracted a good deal of attention.  There are new experimental
results 
\cite{ref1,ref2} 
which motivate the interest.  Furthermore,
there is better theoretical understanding of the process.
The theoretical articles use two distinct methods for their analysis.
The first method considers the production of the pions in the low
$Q^2$ and large $\nu$ region which justifies the application of the 
PCAC relation
\cite{ref3,ref4,ref5,ref6}.
The method is also valid at high neutrino energies, provided the 
above restriction $(\nu \gg \sqrt{Q^2})$ is satisfied.  
The second method 
\cite{ref7,ref8}
studies the coherent production of the $\Delta$--resonance on nuclei
by using a modified $\Delta$--propagator and a
distorted wave--function for the pion.  My article will concentrate
on the first method.  I will state the various conditions entering
the calculation and then remark on the results of recent articles.\\

The first method is based on the following steps.\\

(1)
In neutrino--nucleon or --nucleus scattering there is a kinematic
region where the dominant term of the leptonic current is proportional
to the momentum transfer at the leptonic vertex, denoted by
$q_{\mu}$ 
\cite{ref9}.
This region contains the domain
\begin{equation}
Q^2 = (\rm{a\,\, few}) \cdot m_{\pi}^2
\end{equation}
where the PCAC approximation is valid.  As a result the vector
contribution vanishes (by CVC) and the axial contribution is 
replaced by a pion--nucleus cross section.\\

(2)
Coherent pion production is defined as the process where the four
momentum--transfer squared ($t$) between the current and the produced
pion is small so that the nucleus remains intact.  This was the 
signature of the early bubble chamber experiments -- a sharp peak
in the $t$--dependence.\\

As a consequence of step (1) and the above definition, coherent
pion production by neutrinos is related to the diffractive peak
observed in elastic pion-nucleus scattering.\\

In addition to the dominant term of the leptonic current there are
subdominant contributions, which in the above kinematic region
can be estimated by using data of the reactions $\gamma + N \to
\pi^0 + N$ and $\pi + N \to A + N$. This was done recently in 
\cite{ref5}
where it was established that their contribution is small;
justifying a posteriori the above approximations.

I will give several steps of the calculation in order to appreciate
the accuracy of the result.  We consider the reaction
\begin{equation}
\nu(k_1) N(p) \to \mu^-(k_2) \pi (p_{\pi})N (p')
\end{equation}
where the momenta are indicated in the parentheses.  The invariant
amplitude for the process is
\begin{equation}
T_W = \frac{G}{\sqrt{2}}\, V_{ud} \bar{u}(k_2)\gamma^{\mu}
  (1-\gamma_5) u(k_1)\, \left[ V_{\mu}^+ -A_{\mu}^+\right]\, .
\end{equation}
For coherent scattering there is no vector$\otimes$axial
inference.  The matrix element of the axial current is indicated
by $A_{\mu}^+$ and consists of the pion pole contribution and the 
rest we call $R_{\mu}$.  Therefore
\begin{equation}
-i\, A_{\mu}^+ = \frac{\sqrt{2}f_{\pi}q_{\mu}}{Q^2+m_{\pi}^2}\, T
  (\pi + N\to \pi^+ N) -R_{\mu}\, .
\end{equation}
The PCAC relation now reads
\cite{ref5}
\begin{equation}
q^{\mu} R_{\mu} = -\sqrt{2} f_{\pi} T(\pi^+ N\to \pi^+ N)\, .
\end{equation}
In the domain of coherence the variables attain a simple form.
In the limit $\nu\gg \sqrt{Q^2}$ we use the approximations
\begin{equation}
q_{\mu} = (q_0,\, 0,\, 0,\, \sqrt{\nu^2+Q^2}) \approx
     (\nu,\, 0,\, 0,\, \nu +\frac{1}{2}\, \frac{Q^2}{\nu})
\end{equation}
\begin{equation}
-q^2 = Q^2 = m_{\ell}^2\, \frac{y}{1-y} \, + k_1^0 k_2^0 \theta^2
\quad\quad{\rm with}\quad\quad y =\frac{\nu}{E_1}\, .
\end{equation}
Of particular interest is the experimental configuration $\theta =0$
when the muon is parallel to the neutrino (in the laboratory frame).
In this case
\cite{ref9}
\begin{equation}
Q^2 = m_{\ell}^2\, \frac{y}{1-y},\,\, k_{1\mu} = \frac{1}{y}q_{\mu}
\quad\quad{\rm and}\quad\quad k_{2\mu} = \frac{1-y}{y}q_\mu\, .
\end{equation}
There are also smaller terms in these formulas which have been 
estimated to give smaller cross sections 
\cite{ref5}.\\

The cross section is calculated in a straight--forward way.
After squaring the matrix element $T_W$ we can explicitly calculate
the leptonic tensor in terms of traces.  In the parallel configuration
we make the substitutions in equation (8), which indicate that all
surviving lepton terms are proportional to $q_{\mu}$, as stated in
item (1) at the beginning of this article.  Thus the vector
contribution $V_{\mu}^+$ vanishes, the pion pole is calculated
explicitly and $q^{\mu}R_{\mu}$ is determined by PCAC.  This way
I reproduced the original Adler formula 
\cite{ref9}.
Comparisons with experimental data require an extrapolation of the 
formulas to finite values of $Q^2$, which is mentioned in the 
original article
\cite{ref9}.
Later on an improved extrapolation was introduced by using for
$Q^2$ eq.~(7) instead of (8) in the square of the pion--pole term
\cite{ref10}.
With these steps the original formulas are reproduced and the
approximations are evident.\\

In our work 
\cite{ref5}
we analysed coherent $t$--pion production in terms of helicity cross
sections.  We treat the kinematics exactly by calculating the density
matrix elements.  Among the four polarizations we only approximate
the helicity zero polarization as follows 
\begin{equation}
\in_{\mu}(\lambda=0) = \frac{1}{\sqrt{Q^2}}
  \left(|\vec{q}|,\, 0,\, 0,\, q_0\right) \approx
    \frac{q_{\mu}}{\sqrt{Q^2}} + 0\left(\frac{Q^2}{\nu^2}\right)
\end{equation}
in order to apply the PCAC relation.  The other polarizations with
exact density matrix elements are kept in the calculation.  This is
evident in equations (7), (13) and (16) in ref.\
\cite{ref5}.
For $E_{\nu} < 4$ GeV we use experimental data for either 
calculating or estimating the various terms.  We found that the
helicity zero and the longitudinal cross sections dominate.  Our
formulas look more complicated but they are more accurate because
they include the explicit muon mass, $Q^2$ and $\nu$ dependence of
the density matrix elements.

When we write out the pion pole explicitly, we expect the remaining
hadronic cross sections to be smooth functions of $Q^2$ for $Q^2$
\raisebox{-0.1cm}{$\stackrel{<}{\sim}$} $0.1$ GeV$^2$.  Thus the 
$Q^2$--extrapolation is given by the kinematic factors appearing
in our formulas.  The presence of the muon mass reduces the 
charged current cross section in the small $Q^2$ domain.  This is
evident in the functional form of $\tilde{L}_{\rm oo}$ and
$\tilde{L}_{\ell\rm o}$ in equation (6) of ref.\
\cite{ref5}.
The net effect is a reduction of the differential cross section
$d\sigma/dQ^2$ in the small $Q^2$ (see figure 2 in \cite{ref5}).
A reduction is also reported in a recent article 
\cite{ref6}
where the authors use the improved $Q^2$ extrapolation
\cite{ref10}.

Using our formulas we calculated the differential $d\sigma/dQ^2$
and the integrated cross sections.  We found a substantial reduction
for the $CC$ reaction.  The reduction brings better agreement with the
K2K upper limit.\\

To sum up, coherent pion production by neutrinos represents a very
attractive reaction which is theoretically well understood.  Looking
at the derivation I expect that a careful calculation should give
a theoretical cross section with an accuracy of 20 to 30\%, which
is typical of PCAC predictions.  For this reason we are now 
repeating the low energy ($E_{\nu}<4$ GeV) calculation and plan to
extend it to higher neutrino energies where data is already available.
A successful explanation of all data will indicate that we possess
an accurate theoretical calculation.  Consequently, this reaction
together with quasi--elastic scattering and resonance production
will serve in the future as a bench--mark for determining the 
experimental flux and properties of the neutrinos.


\begin{theacknowledgments}
  I wish to thank Dr. C.~Hill and the theory group for their hospitality
at Fermilab.  I also thank Dr. J.~Morfin for organizing a wonderful
meeting.
\end{theacknowledgments}

\end{document}